\documentclass[11pt,a4paper]{article}
\usepackage[latin1]{inputenc}
\usepackage{amsmath}
\usepackage{amsfonts}
\usepackage{amssymb}
\usepackage[dvipsnames]{xcolor}
\usepackage{mathtools}
\usepackage{jheppub}

\newcommand{\ee}[1]{\begin{equation}#1\end{equation}}
\newcommand{\ea}[1]{\begin{align}#1\end{align}}

\providecommand{\f}[2]{\frac{{#1}}{{#2}}}

\title{Dark energy without fine tuning}

\author[a]{Jos\'e Eliel Camargo-Molina,}
\author[a, b]{Tommi Markkanen}
\author[a]{and Pat Scott}
\emailAdd{j.camargo-molina@imperial.ac.uk}
\emailAdd{t.markkanen@imperial.ac.uk}
\emailAdd{tommi.markkanen@kbfi.ee}
\emailAdd{p.scott@imperial.ac.uk}

\affiliation[a]{Department of Physics, Imperial College London, Blackett Laboratory, London, SW7 2AZ, UK}
\affiliation[b]{National Institute of Chemical Physics and Biophysics, R\"{a}vala 10, 10143 Tallinn, Estonia}

\abstract{\noindent We present a two-field model that realises inflation and the observed density of dark energy today, whilst solving the fine-tuning problems inherent in quintessence models. One field acts as the inflaton, generically driving the other to a saddle-point of the potential, from which it acts as a quintessence field following electroweak symmetry breaking.  The model exhibits essentially no sensitivity to the initial value of the quintessence field, naturally suppresses its interactions with other fields, and automatically endows it with a small effective mass in the late Universe.  The magnitude of dark energy today is fixed by the height of the saddle point in the potential, which is dictated entirely by the scale of electroweak symmetry breaking.}

\begin{document}
\begin{flushleft}
	\hfill		  IMPERIAL/TP/2019/TM/04
\end{flushleft}
\maketitle

\section{Introduction}
\label{sec:intro}
Explaining the accelerated expansion of the Universe \cite{Riess:1998cb,Perlmutter:1998np} with a cosmological constant $\Lambda$ requires an unacceptable amount of fine tuning, due to the extreme smallness of the observed density of dark energy $\rho_\Lambda\sim2.5\cdot10^{-47}$\,GeV$^4$ \cite{Peebles:2002gy,Martin:2012bt,Sola:2013gha,Aghanim:2018eyx}. Here we define fine tuning as the careful adjustment of initial conditions, couplings or differences of couplings, to very large or very small non-zero values. A possible and popular mechanism for alleviating this fine tuning is to construct a theory where the energy of the vacuum is strictly zero, forbidding a $\Lambda$ term in the Lagrangian.  For example, this is often achieved by invoking some additional (often unspecified) symmetry.  The current observation of an approximately constant density of dark energy is then explained by the gradual dynamics of fields \cite{Copeland:2006wr}.

Similar to inflation, many models for dark energy contain slowly-rolling scalar fields \cite{Wetterich:1987fm,Ratra:1987rm}. These are usually referred to as ``quintessence'' models. However, they frequently suffer from issues related to fine tuning and naturalness.  Even if not always the case \cite{Copeland:1997et,Ferreira:1997hj,Zlatev:1998tr,Steinhardt:1999nw}, predictions are often highly sensitive to the initial conditions.  Successful quintessence generally requires very small (i.e.\ fine-tuned) parameters, such as masses of the order of $m\sim10^{-33}$\,eV.  Similarly, couplings of the quintessence field to other fields, which should otherwise be allowed by the known gauge and Lorentz symmetries of the Standard Model (SM), must also be heavily suppressed or screened \cite{Khoury:2003aq,Khoury:2003rn,Brax:2004qh,Hinterbichler:2010es} in order to avoid fifth-force constraints [\citenum{Adelberger:2003zx,Will:2005va,Adelberger:2009zz}; see also recent experimental proposals \citenum{Burrage:2014oza,Sabulsky:2018jma}].

A rather unexpected connection was recently discovered \cite{Dimopoulos:2018eam} between the observed dark energy density $\rho_\Lambda$, cosmic inflation and electroweak symmetry breaking:
\ee{\rho_\Lambda\approx\f{v^8}{{\mathcal P}_\zeta M_{\rm P}^4}\,,\label{eq:conspiracy}}
where ${\mathcal P}_\zeta\approx2.2\cdot10^{-9}$ is the observed amplitude of the spectrum of primordial perturbations at microwave background scales \cite{Ade:2015xua}, $M_{\rm P}$ is the {reduced} Planck mass and $v = 246$\,GeV is the vacuum expectation value (VEV) of the Higgs field $h$. The relation (\ref{eq:conspiracy}) is more than just a curious coincidence between parameters; in many models it arises as the magnitude of the potential energy at electroweak symmetry breaking (EWSB) left over from the interplay of the Higgs boson and the inflaton \cite{Dimopoulos:2018eam}.

Unfortunately, (\ref{eq:conspiracy}) is not a panacea for all fine-tuning problems of quintessence. Even if (\ref{eq:conspiracy}) can generate the scale of dark energy, successful quintessence still requires a small enough effective mass, the suppression of couplings to other fields, and fine-tuned initial conditions. The main result of this work is to show that all these issues can be naturally resolved in a two-field model, whilst simultaneously explaining the observed magnitude of dark energy via (\ref{eq:conspiracy}).  We use positive metric, Riemann and Einstein sign conventions $(+,+,+)$ \cite{Misner:1974qy}.

\section{Dark energy scale from inflation and EWSB}
\label{sec:scale}
The tree-level potential for the Higgs at $T=0$ can be conveniently parameterised as
\ee{V(h)=\f{\lambda}{4}\left(h^2-v^2\right)^2\,,\label{eq:Hig}}
where the total vacuum energy {--  including the zero-point quantum fluctuations --} is assumed to strictly vanish  at $h=v$, in accordance with the discussion in the introduction with the hope that in a complete theory it has a more fundamental motivation. Crucially, the potential energy from the Higgs changes at EWSB. Above the critical temperature $T_{\rm EW}$ at which the mass parameter of the Higgs potential $m^2_H(T_{\rm EW}) = 0$, the potential is minimised at $h = 0$. At $T < T_{\rm EW}$, we see that $m^2_H(T < T_{\rm EW}) < 0$, such that the potential recovers its familiar Mexican-hat shape and $h = 0$ is rendered a local maximum \cite{Arnold:1992rz}
\ee{V(h_{\rm min})=\begin{cases}V(0)=\f{\lambda}{4}v^4\,, &T>T_{\rm EW}\\V(v)=0\,, &T<T_{\rm EW}
	\end{cases}\,.\label{eq:change}}

Consider the following action $S=\int\sqrt{-g} \mathcal{L}$ for the inflaton $\phi$, where
\ea{-\mathcal{L}& = \f{1}{2}\partial_\mu\phi\partial^\mu\phi+U(\phi,h)\label{eq:L}\\
     & = \f{1}{2}\partial_\mu\phi\partial^\mu\phi+\f{1}{2}m^2\phi^2+{c}\f{\phi}{M_{\rm P}}V(h)\,.\nonumber}
In addition to the usual quadratic piece defining the bare inflaton mass $m$, the potential of this theory possesses a linear Planck-suppressed coupling of the inflaton to the Higgs, with strength characterised by the constant $c$.

Before EWSB ($T>T_{\rm EW}$), the Higgs has non-zero vacuum energy $V(0) =(\lambda/4)v^4$, slightly displacing the minimum of the inflaton potential from the origin,
\ee{U'(\phi_0,0)=0\quad \Leftrightarrow\quad \phi_0=- \f{cV(0)}{M_{\rm P}m^2}=-\f{c\lambda v^4}{4M_{\rm P}m^2}\,.\label{eq:min}}
After EWSB ($T<T_{\rm EW}$), once the Higgs reaches its (new) minimum $h=v$, a second rolling of the inflaton is triggered starting from $\phi = \phi_0$.
{Choosing $|c| \approx 1.7 $ and using $M_{\rm P}=2.43 \cdot 10^{18} \, \mbox{GeV}$, $\lambda = 0.129$ and $v = 246 \, \mbox{GeV}$, we see that for the quadratic model of inflation, $m\approx6\cdot10^{-6}M_{\rm P}\sim 10^{-1}\sqrt{{\mathcal P}_\zeta}M_{\rm P}$, the initial value of the potential energy at the beginning of this rolling phase is
\ee{
{U(\phi_0,v)}=\f{c^2\lambda^2 v^8}{32M_{\rm P}^2m^2}\approx  \frac{ (10 \, c \, \lambda \, v^4)^2}{32 M_{\rm P}^4 \mathcal{P_\zeta} } \approx  2.5 \cdot 10^{-47} \, \mbox{GeV}^4\,,\label{eq:vac_e}}
which} agrees with the observed value of $\rho_\Lambda$ \cite{Aghanim:2018eyx}. Dark energy with the magnitude (\ref{eq:conspiracy}) is thus a direct prediction of inflation and electroweak physics in this setup, as long as the potential energy does not change significantly after EWSB. The relation (\ref{eq:vac_e}) is a consequence of the specific linear coupling to the Higgs $\f{\phi}{M_{\rm P}} V(h)$. Here we
tacitly stay within the vicinity of the minimum $\phi_0$, where the linear term dominates and for now assume that problems such as boundedness of the potential are resolved by additional, unspecified, higher-order terms. In Sec.\ \ref{sec:mr} we present a realistic scenario that indeed has no such issues.
The symmetries of the theory also permit a portal term $\phi^2 h^2$,
however such a term has only a very small impact on the end result, as the mass of the inflaton is significantly larger than the electroweak scale. On the other hand, a linear term of the form $\phi M^3$ with a scale $M^3\gtrsim \f{1}{M_{\rm P}} V(0)\sim \f{v^4}{M_{\rm P}}$ would spoil the relation (\ref{eq:vac_e}).

Qualitatively the same features as in (\ref{eq:L}) are present in the Starobinsky model of inflation \cite{Starobinsky:1979ty}, with the added bonus that a coupling between the Higgs and the inflaton is generated automatically \cite{Dimopoulos:2018eam}.

Unfortunately, even if the dark energy density $\rho_\Lambda$ somewhat miraculously arises as a function of known and observable scales, the inflaton remains much too heavy to act as a quintessence field. For it to roll slowly today, its mass must be much smaller than the current Hubble rate, an extremely small number
$H_0\sim {\sqrt{\rho_\Lambda}}/{M_{\rm P}}\sim 10^{-33}$eV.
Fortunately, there are ways around this problem.

\section{Slow-roll from a non-canonical kinetic term}
\label{sec:sr}
The general idea of ameliorating the problems of quintessence with non-canonical kinetic terms has existed for some time \cite[see e.g.][]{Chiba:1999ka,Hebecker:2000au,Hebecker:2000zb,ArmendarizPicon:2000ah,ArmendarizPicon:2000dh}.
In this vein, let us now modify the theory characterised by the action (\ref{eq:L}) by taking the kinetic term to have a non-canonical form
\ee{(\partial_\mu \phi)^2\quad\longrightarrow\quad\bigg(\f{M_{\rm P}}{\phi}\bigg)^2(\partial_\mu \phi)^2\equiv (\partial_\mu \chi)^2\,,\label{eq:alpha}}
with the same notation as introduced in the Appendix. {In the minimum defined by (\ref{eq:min}), the potential after EWSB now has precisely the behaviour required for quintessence of the thawing variety \cite{Caldwell:2005tm}, $\tilde{U}(\chi_0,v)\sim\rho_\Lambda$ and $\tilde{U}''(\chi_0,v)\sim H_0^2$.} By introducing an $O(1)$ coupling in front of the non-canonical kinetic term (\ref{eq:alpha}), one may evade current observational bounds on the parameter of state for dark energy \cite{Dimopoulos:2018eam}.  We explore these bounds quantitatively for our two-field model in Sec.\ \ref{sec:ar}.

The modification (\ref{eq:alpha}) not only makes the mass of $\chi$ effectively small (see Appendix), but also suppresses all interactions with other fields. Consider, for example, a Yukawa coupling of the form $g\phi\bar{\psi}\psi$. Close to the minimum (\ref{eq:min}) in the canonical variable  $\chi$, the effective Yukawa coupling is $\tilde{g}\sim g\sqrt{\rho_\Lambda/(m^2 M_{\rm P}^2)}$, which leads to a negligible interaction strength. This is very similar to the suppression of interactions that occurs in $\alpha$-attractor models of inflation \cite{Kallosh:2016gqp,Dimopoulos:2017tud}.

Unfortunately, the term (\ref{eq:alpha}) also makes it extremely difficult for the inflaton field to reach the minimum (\ref{eq:min}) in the first place. In the early Universe, a very light field 
will be stopped by Hubble friction long before ever reaching such a minimum. Decay via interactions (required for reheating) is also virtually impossible, due to the manifest suppression of all interactions, as in our earlier Yukawa example. This is not a problem unique to our scenario, but rather a common feature of quintessence models (albeit not all such models, as discussed in Sec.\ \ref{sec:intro}).  Avoiding this usually requires very careful fine-tuning of the initial field conditions. 

To avoid this issue more naturally, we need a mechanism that first allows the field to reach the minimum unhindered, and only then triggers the non-canonical kinetic behaviour (\ref{eq:alpha}). In Ref.\ \cite{Dimopoulos:2018eam}, this was dubbed the {\it bait-and-switch} mechanism, and several examples were also provided. The most natural one comes by coupling (\ref{eq:alpha}) with the Higgs, such that the kinetic term is canonical up until EWSB, providing ample time for the relaxation into the minimum. Although not problematic at the classical/mean-field level, coupling the kinetic term to the Higgs introduces interactions that are likely already excluded by collider bounds (on e.g.\ invisible Higgs decays). The other examples provided in Ref.\ \cite{Dimopoulos:2018eam} were more for illustrative purposes, and not expected to be easily realised in top-down approaches. In Ref.\ \cite{Wang:2018kly}, a mechanism designed to extend that of Ref.\ \cite{Dimopoulos:2018eam} was presented, requiring a very large non-minimal coupling between the Higgs and the scalar curvature of gravity, of the order $|\xi|\sim 10^{32}$.

\section{A two-field model: assisted relaxation}
\label{sec:ar}
So far we have only discussed quintessential inflation \cite{Peebles:1998qn}, where inflation and dark energy are given by the same field $\phi$. The mechanism that we now present relies on the relation (\ref{eq:conspiracy}) and the stretching of the potential by a non-canonical kinetic term of the form (\ref{eq:alpha}), but with the crucial difference that it includes a second field in addition to the inflaton. It is this additional field that sources dark energy in the late Universe, in contrast to 
Ref.\ \cite{Dimopoulos:2018eam}. What we will show is that the dynamics of the inflaton can assist and effectively force the relaxation of a field with a flat potential into the pre-EWSB minimum.  This provides a natural resolution of the initial-value problem, leading to a model that does not suffer from {\it any} of the usual fine-tuning issues of quintessence.

Models involving multiple quintessence fields have been previously presented in the literature; see for example \cite{Barreiro:1999zs,Fujii:1999fc,Masiero:1999sq,Blais:2004vt,Kim:2005ne,Elizalde:2008yf,vandeBruck:2009gp,Akrami:2017cir}.
Many-field models with a connection to Early Universe inflation, as in our set-up, are less common (although not absent \cite{Akrami:2017cir,Elizalde:2008yf}), as are embeddings in more fundamental theories. However, although our study also does not incorporate a fundamental theoretical framework, the novel connection between electroweak theory and inflation potentially provides a clue that can aid in finding top-down manifestations of our mechanism.

Our mechanism likely has many manifestations. We will first focus on the simple Lagrangian $\mathcal{L} = \mathcal{L}_\phi + \mathcal{L}_\sigma$, where $\phi$ is the inflaton, $\sigma$ the quintessence field, $-\mathcal{L}_\phi \equiv \f12\partial_\mu\phi\partial^\mu\phi +\f12m^2\phi^2$ and
\ea{-\mathcal{L}_\sigma& = b^2\f{1}{2}\f{M_{\rm P}^2}{\phi^2+\sigma^2}\partial_\mu\sigma\partial^\mu\sigma + \f{1}{2}m^2\sigma^2+{c}\f{\sigma}{M_{\rm P}}V(h) \nonumber \\
                & \equiv \f12 \partial_\mu\chi\partial^\mu\chi+\tilde{U}_\sigma(\chi,h)\,.\label{2field}}
The constants $b$ and $c$ are $\mathcal{O}(1)$ dimensionless couplings. Crucially, we assume that $\sigma$ couples to the Higgs similarly to (\ref{eq:L}), letting it acquire a non-zero VEV before EWSB.  For simplicity, we choose the same mass $m$ for $\phi$ and $\sigma$, as we expect that both parameters are sourced from similar physics --- but their exact masses can differ without introducing the need for any fine tuning. Note also that our mechanism does not rely on the quadratic form for the inflaton potential (which we have also chosen for simplicity), and works equally well for any other theory of inflation.

Non-canonical kinetic terms can be interpreted as a metric $G$ in field space, i.e. $\mathcal{L} \subset G^{IJ} \Phi_I \Phi_J$, with $\Phi$ the fields ($\Phi = [\phi,\sigma]$ in our case). As such, the signature of $G$ cannot be modified by a change of coordinates (in this case field redefinitions), provided the field-space metric is invertible.  Throughout, we will consider models where only one of the fields possesses a non-canonical kinetic term. This means that $G$ is always diagonal.  If the non-canonical kinetic term is always positive (as it is in all our examples), it then follows that there are no ghosts in the models that we present.

If $\phi > M_{\rm P}$ during inflation, the non-canonical kinetic term in (\ref{2field}) effectively makes $\sigma$ heavy enough to roll, even when close to $\sigma=0$ (see Appendix). For a large range of initial conditions, if inflation lasts long enough $\sigma$ will therefore be driven towards its minimum by the inflaton. We call this {\it assisted relaxation}.

When inflation has ended and the Universe has reheated, $\phi=0$, the same kinetic term makes $\sigma$ exponentially light.  This allows $\sigma$ to source dark energy precisely as discussed in Sec.\ \ref{sec:sr}. As shown in Sec.\ \ref{sec:scale}, a linear coupling to the Higgs potential and a tree-level mass of the order required for successful inflation ($m\sim10^{-6}M_{\rm P}$) leads to an excellent match to observations. Indeed, this is the mass scale that should be chosen for $\sigma$ in order to avoid spoiling (\ref{eq:conspiracy}), regardless of actual inflaton potential.

Let us explicitly show the mechanism at work for the model (\ref{2field}). Deep in inflation, $\phi\gg M_{\rm P}$, and (neglecting the derivatives of $\phi$ in the change of variables, as they are subleading in the slow-roll expansion)
\ea{-\mathcal{L}_\sigma & \approx \f{1}{2}\bigg(\f{b M_{\rm P}}{\phi}\bigg)^2\partial_\mu\sigma\partial^\mu\sigma + \f{1}{2}m^2 \sigma^2+c\f{\sigma}{M_{\rm P}}V(0) \label{2field2} \\\label{2field30}
                 & \equiv \f{1}{2}\partial_\mu\chi\partial^\mu\chi + \frac12 \f{m^2}{(bM_{\rm P})^2}\phi^2\chi^2+c\f{\lambda v^4}{4bM_{\rm P}^2}\phi\chi\,,}
where $\sigma \sim \phi\chi/(bM_{\rm P})$. An important assumption in (\ref{2field2}) is that the Higgs is at $h=0$. This is not particularly constraining, as it can be arranged via a portal coupling to the inflaton or a non-minimal coupling to curvature.

Solving the equation of motion for $\sigma$ (or $\chi$) from (\ref{2field2}) whilst holding $H$ and $\phi$ constant, we see that the average energy density of the quintessence field, $\rho_\sigma\sim \f12 m^2\sigma^2$, dilutes approximately as either $\rho\propto e^{-3N}$ or $\rho\propto e^{-\f{4}{b^2}N}$, where $N=H t$ is the number of $e$-folds since the beginning of inflation.  Here we have made the approximation that the energy density is completely dominated by the inflaton. The first case corresponds to a heavy field oscillating around its minimum, i.e.\ acting as normal matter \cite[see e.g.][]{Turner:1983he}.  The second case corresponds to a light field slow-rolling down its potential.  The two cases are approximately distinguished by whether or not the slow roll parameter
\ee{\eta_\sigma=\f{\tilde{U}_\sigma''(\chi,0)}{3H^2}=\f{\f{\phi^2}{(bM_{\rm P})^2}m^2}{{\f{1}{2}\f{m^2}{M_{\rm P}^2}\phi^2}}=\f{2}{b^2}\,}
is larger (leading to $\rho\propto e^{-3N}$) or smaller than unity (leading to $\rho\propto e^{-\f{4}{b^2}N}$).

As a representative initial condition, consider $\sigma=M_{\rm P}$.  Setting $b=1$, leading to a coherently oscillating $\sigma$, we have the approximate behaviour $\rho \propto \sigma^2 \propto e^{-3N}$, so { $\sigma$ approaches the minimum $\sigma_0$  as
\ea{\sigma - \sigma_0 &= \sigma + \f{c\lambda v^4}{4M_{\rm P}m^2}
=\left(M_{\rm P}+\f{c\lambda v^4}{4M_{\rm P}m^2}\right) e^{-\f{3N}{2}}\nonumber \\&\approx M_{\rm P}e^{-\f{3N}{2}}\,.}}
This translates into a minimal required duration of inflation for our mechanism to work, as inflation must continue long enough for the quintessence field to relax into the minimum.  In terms of $e$-folds, this is
\ee{e^{-\f{3N}{2}}\lesssim \f{|c|\lambda v^4}{4M_{\rm P}^2m^2}\sim \f{\sqrt{\rho_\Lambda}}{M_{\rm P}m}\Rightarrow {\rm for}\ |c| = 1,\ N \gtrsim {84}\,. \label{eq:efolds}}
Importantly, this number is not very sensitive to the initial condition for $\sigma$ in units of $M_{\rm P}$. Observations require only that inflation extends over at least 50--60 e-folds, and in many theories it continues for much longer.  Our model is therefore naturally consistent with both theoretical expectations and observational bounds on $N$.  Inflation is therefore generically expected to drive the $\sigma$ field to its minimum, for a wide range of initial conditions. Note also that although we have assumed an initial hierarchy $\phi\gg\sigma$, this can be expected to arise more or less automatically: as visible in (\ref{2field2}), the effective mass for $\chi$ is generally larger than $m$ deep into inflation, making $\sigma$ roll faster than $\phi$ and hence quite generically leading to $\phi\gg\sigma$. We chose not to include a portal term $\sim \phi^2\sigma^2$ in (\ref{2field}), however such a term can be added: when $\phi^2\sigma^2\gtrsim m^2\sigma^2$, it pushes the minimum for $\sigma$ closer to the origin, from which $\sigma$ rolls towards $\sigma_0$ when $\sigma_0\lesssim\phi\lesssim m$.

Suppose $b^2=10$ instead, corresponding to the case of a light, slowly-rolling field. The left-hand side of (\ref{eq:efolds}) then becomes $e^{-\f15 N}$, leading to the requirement that $N\gtrsim {633}$. Again, this is not an unrealistic number.

Once the inflaton has decayed, $\sigma$ becomes exponentially light due to the pole $\sim\sigma^{-2}$ in the kinetic term and is frozen at $\sigma_0$, unaffected by the thermalization of the Higgs. After EWSB, when the Higgs has rolled to its minimum,
\ea{-\mathcal{L}_\sigma & \approx \f12\f{b^2 M_{\rm P}^2}{\sigma^2}\partial_\mu\sigma\partial^\mu\sigma+\f12m^2 \sigma^2\\
    & \equiv \f12\partial_\mu\chi\partial^\mu\chi + \f12{m^2}{M_{\rm P}^2}e^{\f{2\chi}{b M_{\rm P}}}\,,\label{2field3}}

with the initial condition at EWSB
\ee{\sigma_0=\pm M_{\rm P}e^{\f{\chi_0}{b M_{\rm P}}}=-\f{c\lambda v^4}{4M_{\rm P}m^2}\,,\label{eq:chi0} }
where ``$\pm$'' refers separately to the cases where $c>0$ ($-$) or $c<0$ ($+$).  Choosing $|c|\approx 2.1$ and $m\approx 6 \cdot10^{-6}M_{\rm P}$, and again that $M_{\rm P}=2.43 \cdot 10^{18} \, \mbox{GeV}$, $\lambda = 0.129$ and $v = 246 \,$GeV, this gives
\ee{\rho_\Lambda=\tilde{U}_\sigma(\chi_0,v)=\f12m^2\sigma_0^2=\f{c^2\lambda^2 v^8}{32M_{\rm P}^2m^2}\approx2.5\cdot10^{-47}{\rm GeV}^4\,.}

At early times, to a very good approximation $w=-1$ and $\rho_\Lambda$ is constant.  The present-day deviation from $w=-1$ can be parameterised in standard fashion as
\ee{w \equiv -1 + \delta w = \f{X^2 - 6}{X^2 + 6}\,;\quad X \equiv M_{\rm P}\f{\tilde{U}'_\sigma(\chi,v)}{\tilde{U}_\sigma(\chi,v)}=\f{2}{b}\,.}
Taking e.g.\ $b = 1$ leads to $w = -0.2$, and $b^2 = 10$ gives $w = -0.875$.  The current observational bound of $\delta w \lesssim 0.12$ at 90\% confidence \cite{Abbott:2018wog} means that our scenario is consistent with existing constraints for $b\gtrsim 3.2$.

Light fields on an inflating background are known to generate large quantum fluctuations $\delta \chi$, which for a field with a mass $M$ can be shown to saturate with the root-mean-square fluctuation $\sqrt{\langle\delta \chi^2\rangle} \sim H^2/M$.  On the other hand, massless fields have fluctuations that grow without bound \cite{Starobinsky:1994bd} as $\sqrt{\langle\delta \chi^2\rangle} \sim \sqrt{N} H$. From (\ref{eq:chi0}) we see that the minimum in the canonical variable occurs at $\chi_0\lesssim -100 M_{\rm P}$, meaning that quantum fluctuations are generally not large enough to significantly displace the field: the effective mass when $\phi \gtrsim M_{\rm P}$ can be read from (\ref{2field30}), giving $\sqrt{\langle\delta \chi^2\rangle}\sim m\phi/M_{\rm P}$.  Furthermore, inflation is not expected to last for more than a few $e$-folds in the field regime where the effective mass is small ($\phi \lesssim M_{\rm P}$).

\section{A more realistic model}
\label{sec:mr}
When writing (\ref{2field}), we simply assumed that $\sigma$ is coupled to the SM only through a very particular Planck-suppressed linear coupling to the Higgs potential. From a model-building point of view, however, this interaction is also somewhat non-trivial to achieve. Specifically, similar couplings to the inflaton linear in $\sigma$ will likely spoil the mechanism. In this section we will discuss a model where the coupling structure is better justified. The model that we will present here also possesses a symmetry in the UV limit between the inflaton $\phi$ and quintessence field $\sigma$, giving further motivation for the idea that $\phi$ and $\sigma$ may have similar masses. { It also possesses a more realistic non-canonical kinetic term, as we will elaborate in Sec.\ \ref{sec:conc}.}

Let us postulate the following Lagrangian
\ea{&-{\mathcal L} = -\f12 M_{\rm P}^2 e^{c\f{\sigma}{M_{\rm P}}}R+\f{3}{4}c^2e^{c\f{\sigma}{M_{\rm P}}}\partial_\mu\phi\partial^\mu\phi \label{eq:promo} \\
    &          + \f12 b^2 e^{-\phi/M_{\rm P}}\left( e^{-\f{\sigma}{M_{\rm P}}}-1\right)^{-2}\partial_\mu\sigma\partial^\mu\sigma +V(h)+\cdots \nonumber \\
    &          +\alpha^2M_{\rm P}^4\left[\left( e^{-\f{\phi}{M_{\rm P}}}-1\right)^2+\left( e^{-\f{\sigma}{M_{\rm P}}}-1\right)^2\right]\left(e^{c\f{\sigma}{M_{\rm P}}}\right)^2\,,\nonumber}
where the dots signify all SM contributions not written explicitly, { which we drop from now on}. Here we choose $\alpha$ such that $\phi$ leads to successful inflation, and $b$ and $c$ are dimensionless couplings similar to those in (\ref{2field}). { Terms similar to the first one are often present in those scalar-tensor theories where the Planck mass is promoted to a function of a field, e.g.\ \cite{Clifton:2011jh} and will introduce a linear coupling to the Higgs potential, which is manifest in the Einstein frame.} Denoting Einstein-frame quantities with an overline, {we use the standard relations \footnote{Here we drop a $\Box$ term as the space is  unbounded.},
\begin{equation}
g_{\mu\nu}=\Omega^{-2}\overline{g}_{\mu\nu}~~\Rightarrow
~~ R/\Omega^2=\overline{R}-\f{3}{2}\overline{g}^{\mu\nu}\partial_\mu(\ln \Omega^2)\partial_\nu(\ln \Omega^2),\nonumber
\end{equation}
with $\Omega^2=e^{c\f{\sigma}{M_{\rm P}}}$, to write  (dropping the overlines)
\ea{&-{\mathcal L}  = -\f12 M_{\rm P}^2 R+\f34 c^2 \Big(\partial_\mu\phi\partial^\mu\phi + \partial_\mu\sigma\partial^\mu\sigma \Big) \label{eq:promo0}\\
      &
      + \f12 b^2 e^{-\f{\phi}{M_{\rm P}}}\left(e^{-\f{\sigma}{M_{\rm P}}}-1\right)^{-2}e^{-c\f{\sigma}{M_{\rm P}}}\partial_\mu\sigma\partial^\mu\sigma \nonumber \\
      &
      + \alpha^2M_{\rm P}^4\left[\left( e^{-\f{\phi}{M_{\rm P}}}-1\right)^2+\left( e^{-\f{\sigma}{M_{\rm P}}}-1\right)^2\right]
      + e^{-\f{2c\sigma}{M_{\rm P}}}V(h).
       \nonumber }
When} $\sigma\lesssim M_{\rm P}$ the last term introduces a linear coupling to the Higgs potential as required. The term $(e^{c\frac{\sigma}{M_{\rm P}}})^2$ multiplying the last line of \eqref{eq:promo}, automatically avoids couplings between $\phi$ and $\sigma$ in the Einstein frame. We do this for simplicity, as in that case the mechanism follows exactly as in our first example in Sec.\ \ref{sec:ar}. Such couplings do not necessarily spoil the mechanism, but a detailed analysis would be necessary for specific realisations.

\textit{The limit $\phi\gg M_{\rm P},\sigma\gg M_{\rm P}$.}
In this limit the theory (\ref{eq:promo0}) simplifies approximately to
\ea{-{\mathcal L} & \approx-\f12 M_{\rm P}^2 R+\f34 c^2 \left(\partial_\mu\phi\partial^\mu\phi+\partial_\mu\sigma\partial^\mu\sigma\right)\label{eq:rmod2}\\
              & \phantom{=} + \alpha^2M_{\rm P}^4\left[\left( e^{-\f{\phi}{M_{\rm P}}}-1\right)^2+\left( e^{-\f{\sigma}{M_{\rm P}}}-1\right)^2\right]\,,\nonumber}
so in this region we have inflation from both fields. We take the $\phi\leftrightarrow\sigma$ symmetry in this limit as motivation for choosing the mass scale $\alpha M_{\rm P}$ to be the same for both fields.

\textit{The regime $\phi\gg M_{\rm P}, \sigma\lesssim M_{\rm P}$.} In this case,
\ea{-{\mathcal L}&\approx-\f12 M_{\rm P}^2 R+\f34 c^2 \left(\partial_\mu\phi\partial^\mu\phi+\partial_\mu\sigma\partial^\mu\sigma\right) +\alpha^2M_{\rm P}^2\sigma^2\nonumber \\
             & \phantom{=} + \alpha^2M_{\rm P}^4\left( e^{-\f{\phi}{M_{\rm P}}}-1\right)^2
             +\left({1}-c\f{2\sigma}{M_{\rm P}}\right)\f{\lambda v^4}{4}
             \label{eq:rmod3}\,.}
In this regime, we have Starobinsky-type single-field inflation from $\phi$, which successfully matches observations for $\alpha\sim 10^{-5}$. Furthermore, the $\sigma$ field will be driven towards the minimum at $\sigma_0= c\lambda v^4/(4\alpha^2 M_{\rm P}^3)$.  The condition for slow-roll of $\sigma$ is $\eta_\sigma \ll 1$.  Taking the canonically-normalised field $\chi = \f{\sqrt{3}}{2}c\sigma$, we see that $\eta_\sigma =\f{8}{3 c^2}$
or that the field rolls slowly only for $c\gtrsim 1$.

For (\ref{eq:rmod3}) to be valid, the pre-factor in front of the quintessence field's kinetic term $(\partial\sigma)^2$ in the {second} line of (\ref{eq:promo0}) must not grow large and flatten the potential when $\sigma\rightarrow\sigma_0$. We can use this to place a crude lower bound on the initial value of $\phi$:
{
\ea{
\f12 b^2 e^{-\f{\phi}{M_{\rm P}}}\left(e^{-\f{\sigma_0}{M_{\rm P}}}-1\right)^{-2}e^{-c\f{\sigma_0}{M_{\rm P}}}&\approx e^{-\f{\phi}{M_{\rm P}}}\f{8\alpha^4 M_{\rm P}^8}{c^2\lambda^2 v^8}
\lesssim1 \nonumber\\
\implies \phi \gtrsim \ln\left(\f{M_{\rm P}^4}{4\rho_\Lambda}\right) M_{\rm P} &\sim 276\, M_{\rm P}\,.\label{eq:phibo}}}
We see that (unlike in the scenario in Sec.\ \ref{sec:ar}), we require a mild hierarchy amongst the initial field values for the model to work, with $\phi\sim{\mathcal O}(100)M_{\rm P}$ for $\sigma\sim M_{\rm P}$.

\textit{The regime $\phi= 0, \sigma\ll M_{\rm P}$.} After the inflaton decays, the $\sigma$-dependent terms are{
\ee{-{\mathcal L}_\sigma  \approx
\f12  \left(\f{b M_{\rm P}}{\sigma}\right)^2\partial_\mu\sigma\partial^\mu\sigma+ \alpha^2M_{\rm P}^2\sigma^2-\f{2c\sigma}{M_{\rm P}}V(h)
        \label{eq:rmod4}\,,}
which} leads to qualitatively identical behaviour to (\ref{2field3}) in the late Universe: the quintessence field will start slow-rolling from $\sigma_0$ following EWSB, sourcing dark energy in the late Universe in a manner consistent with observations if $b$ and $c$ are set to appropriate $\mathcal{O}(1)$ numbers.

Equation (\ref{eq:phibo}) indicates a very long period of inflation with a practically massless field. Repeating the steps laid out at the end of Sec.\ \ref{sec:ar} shows that quantum fluctuations larger than the classical displacement are in fact generated in this model. However, they can be avoided by taking $A$ to be an $\mathcal{O}(1)$ number and making the simple modification $\exp(-\phi/M_{\rm P}) \to \exp(-A\phi/M_{\rm P})$ in the second line of (\ref{eq:promo}).  All the arguments given in this section remain unchanged under this replacement. Taking e.g.\ $c=1$ and $H=10^{11}$\,GeV, this modification lowers the hierarchy in (\ref{eq:phibo}) enough to avoid large quantum corrections when $A\gtrsim5$.  Finally, it is worth noting that even \textit{were} large corrections generated during inflation, their only impact would be to shift $\sigma_0$, the effective value that the field is driven to during inflation.  In that case, after inflation as the Universe cools and the ratio of the fluctuation correlation length to horizon scale decreases, by the time of EWSB the field will simply have relaxed gradually to the same minimum $\sigma_0$ as we have assumed in this section, changing none of the features of the model.

\section{Conclusions}
\label{sec:conc} The mechanism that we present successfully explains the strength of dark energy, via the inflation-assisted relaxation of a quintessence field and electroweak symmetry breaking.
It does this without the fine tuning that traditionally dogs quintessence: the need for a small effective mass, the initial value problem, and the need to forbid interactions of the quintessence field with SM fields. Our mechanism therefore poses significant interest for model building. We also showed that the critical aspects of the theory are expected to be achievable in well-motivated top-down constructions. In particular, the form of the kinetic term that we propose in (\ref{eq:promo}) is likely more realistic than the one postulated in (\ref{2field}). Indeed, theories beyond Einstein gravity can be generically parameterised in the form \cite{Maeda:1988ab,DiMarco:2002eb}
\ee{\mathcal{L}_{\rm GE}=\f{M_{\rm P}^2}{2}R-\f{1}{2}\partial_\mu\phi\partial^\mu\phi-e^{-2F(\phi)}\f{1}{2}\partial_\mu\sigma\partial^\mu\sigma-U(\phi,\sigma)\,.}
For the appropriate choice of $F(\phi)$ and potential, this clearly leads to the same qualitative features as (\ref{eq:promo0}). Finding such a UV-complete theory that realises our mechanism would be of substantial interest, and therefore should constitute a high priority for future work.

\begin{acknowledgments}
{
We are grateful to Andrew Tolley and Matthew Roberts for helpful discussions, and to STFC (ST/K00414X/1, ST/N000838/1, ST/P000762/1) and the Estonian Research Council (Mobilitas Plus MOBJD323) for funding support.
}
\end{acknowledgments}

\bibliography{Quint.bib}

\appendix

\phantom{skip}

\noindent\textbf{Appendix: Non-canonical kinetic terms.}
Consider a Lagrangian with a non-canonical kinetic term
\ea{
      -\mathcal{L}=\f{1}{2}C^2 \partial_\mu\phi\partial^\mu\phi+U(\phi)\,;\qquad U(\phi)=\f{1}{2}m^2\,,}
where $C$ is a number. We can define a canonically normalised field variable $\chi$ simply by demanding
\ee{C^2 \partial_\mu\phi\partial^\mu\phi\equiv\partial_\mu\chi\partial^\mu\chi\quad \Rightarrow\quad \chi = C\phi\,,}
leading to
\ea{
      -\mathcal{L}=\f{1}{2} \partial_\mu\chi\partial^\mu\chi+\tilde{U}(\chi)\,;\qquad \tilde{U}(\chi)=\f{1}{2}\f{m^2}{C^2}\,.}
There are two crucial features arising from having a non-canonical kinetic term: 1) the value of the potential at some given $\phi_0=\chi_0/C$ is the same in both coordinates
\ee{U(\phi_0)=\f{1}{2}m^2\phi_0^2=\f{1}{2}\f{m^2}{C^2}{{C^2}\phi_0^2}=\tilde{U}(\chi_0)\,.}
2) the slope and specifically the effective mass changes
\ee{U''(\phi_0)=m^2\,,\qquad\tilde{U}''(\chi_0)=\f{m^2}{C^2}\,.}
In a nutshell, factors larger than ${\mathcal O}(1)$ in front of kinetic terms make the field lighter and smaller than ${\mathcal O}(1)$ make it heavier.

These features persist also when the kinetic term is multiplied by a function. { An illustrative example (for $\phi >0$) is
\ea{-\mathcal{L} &=\f12 \bigg(\f{M_{\rm P}^2}{\phi^2}\bigg)^2\partial_\mu\phi\partial^\mu\phi+\f12 m^2\phi^2\,;~~ \Rightarrow ~~\phi =  M_{\rm P}e^{\f{\chi}{M_{\rm P}}}\,,\nonumber\\&\Rightarrow ~~U''(\chi)  =\f{4}{M_{\rm P}^2}\tilde{U}(\chi)=2m^2 e^{2\f{\chi}{M_{\rm P}}}\,. }
}So clearly, when $\phi\ll M_{\rm P}$ $(\chi \ll 0)$ the field is very light, whereas for $\phi\gg M_{\rm P}$  $(\chi \gg 0)$ the field is heavy. When the kinetic prefactor is one, $\phi= M_{\rm P}$ $(\chi = 0)$, there is virtually no change in effective mass.

\end{document}